\documentclass[%
 reprint,
 amsmath,amssymb,
 aps,
]{revtex4-2}

\usepackage{graphicx}% Include figure files
\usepackage{dcolumn}% Align table columns on decimal point
\usepackage{bm}% bold math
\usepackage{subfig}

\usepackage{caption}
\allowdisplaybreaks

\begin{document}

\preprint{APS/123-QED}

\title{The Impact of Meta-Strategy on Attendance Dynamics \\ in the El Farol Bar Problem}

\author{Rebecca E. Cohen}
\email{rebecca.cohen@colorado.edu}
\affiliation{%
 Department of Applied Mathematics, University of Colorado at Boulder, Colorado 80309, USA
}%
\author{Juan G. Restrepo}%
 \email{juanga@colorado.edu}
\affiliation{%
 Department of Applied Mathematics, University of Colorado at Boulder, Colorado 80309, USA
}%

\begin{abstract}
The El Farol Bar Problem is a classic computational economics problem in which agents attempt to attend a weekly event at a bar only if it is not too crowded.  Each agent has access to multiple competing strategies that may be used to predict whether attendance will be above the tolerance threshold.  Many different variations of the El Farol Bar Problem have been published, with a particularly broad variation in the choice of \emph{meta-strategy}, the algorithm by which each agent selects the best-performing strategy. This paper discusses the varying mechanisms of self-organization within the system when different meta-strategies are used, including the existence and location of fixed points in the limit of an infinite number of agents.  Informed by these mechanisms, we present an evolutionary paradigm that reduces the likelihood of degenerate cases where attendance is always too high or too low, while reducing the amplitude of fluctuations in attendance.  This evolutionary paradigm sheds light on the critical role of heterogeneity of strategies selected in the emergence of a stable steady state in the El Farol Bar Problem.
\end{abstract}

%\keywords{Suggested keywords}%Use showkeys class option if keyword
                              %display desired
\maketitle

%\tableofcontents

\section{Introduction}

The El Farol Bar Problem was introduced by economist Brian Arthur in 1994 to explore the role of bounded rationality in multi-agent systems \cite{Arthur94}.  In the model, each agent competes to utilize a limited resource.  The scenario presented is a weekly event at a bar with live music that is only enjoyable when not too crowded, but the same concept has been applied to scenarios ranging from traffic flow to stock trading \cite{zambrano2004}.  Each agent must make decisions based on imperfect models of the system, and these decisions determine the future dynamics of the system.

In Arthur's original setup, each agent has a fixed collection of strategies available to them.  At each iteration, the agents select whichever strategy would have performed best in prior weeks and use that strategy to make a decision about whether to attend the event.  In simulations, attendance at the bar often appears to oscillate around the threshold of crowding that agents are willing to tolerate. Many different variations of the problem have been described, most of which converge to a similar state in which attendance fluctuates close to the threshold \cite{Miklos2012, chen2016, Chen2017, zambrano2004,  Challet2004}.  However, at least in some formulations of the problem, the mean attendance is observed to differ from the threshold of crowding that agents are willing to tolerate \cite{RePEc:eee:phsmap:v:258:y:1998:i:1:p:230-236, LUS2005651}, and volatility in attendance dynamics has also been shown to be sensitive to parameters governing the system's behavior \cite{RePEc:eee:phsmap:v:258:y:1998:i:1:p:230-236, LUS2005651, cara1999, RePEc:eee:phsmap:v:269:y:1999:i:1:p:1-8}.

In the El Farol Bar Problem, each agent must make a choice about whether to attend the bar based on models accessible to the agent that do not accurately represent the dynamics of the system.  Previous works are based on two main approaches.  In the first approach, used in the original formulation of the El Farol Bar Problem, each agent has access to a pool of strategies, out of which they select the one that would have performed best during some prior period \cite{Arthur94, chen2016, Chen2017, zambrano2004, Challet2004, stluce_2020, CHALLET1997407, RePEc:eee:phsmap:v:258:y:1998:i:1:p:230-236, RePEc:eee:phsmap:v:269:y:1999:i:1:p:1-8}.  Alternatively, agents may update their strategies based on past behavior of the system \cite{Miklos2012, chen2016, Chen2017, Fogel99, Whitehead2008}.  These two approaches may be used separately or in tandem.  The literature contains many different approaches to updating strategies.  Depending on the implementation, attendance may coalesce to a stable level close to the threshold \cite{Miklos2012, Chen2017}, or fluctuate wildly \cite{Fogel99}.  Some implementations that involve agent adaptations do so by allowing agents to either collaborate or learn from each other's choices.  Intriguingly, allowing agents to collaborate or copy each other's previously effective strategies may lead to greater volatility and lower overall resource utilization  \cite{Chen2017, collins2017}.

Models that allow agents to select between multiple strategies must employ a meta-strategy, that is, an algorithm by which agents select their preferred strategy out of those options available.  There is considerable variation in the literature with regard to how strategies are evaluated.  For example, the accuracy of a strategy may be computed as a binary function of whether the strategy would lead to an agent making the correct choice at each time step \cite{stluce_2020, Challet2004, CHALLET1997407, RePEc:eee:phsmap:v:258:y:1998:i:1:p:230-236} or as a numerical-valued function of the prediction error \cite{cara1999, RePEc:eee:phsmap:v:258:y:1998:i:1:p:230-236, zambrano2004}.

In this paper, we explicitly consider the impact of \emph{meta-strategy}, the algorithm by which an agent selects their strategy, on the dynamics of the system.  We present one meta-strategy that robustly drives oscillations around the threshold but is prone to large fluctuations in many regimes, and another one in which attendance may reach a fixed point either above or below the threshold in certain regimes.  By analyzing the behavior of agents probabilitistically, we find analytical expressions and necessary conditions for the existence of fixed points of the dynamics of attendance.  Furthermore, we present an evolutionary paradigm that improves the robustness of both meta-strategies, leading to smaller fluctuations close to the threshold.

Our paper is organized as follows.  In Section II, we present our model for the El Farol Bar Problem and introduce two different meta-strategies.  In Section III, we analyze the system under each meta-strategy.  In Section IV, we present an evolutionary paradigm and show how it modifies the dynamics of the system.  Finally, we present our conclusions and discuss our results in Section V.

\section{Model}

In this section we specify our version of the El Farol Bar Problem.  We begin with a population of $A$ agents, all of whom prefer to attend the bar only if the number of agents attending will be less than a threshold $T$.

Each agent has a choice of $s$ different strategies, out of which they pick the one they expect to perform best, according to a fixed meta-strategy.  At time $t$, we denote the previous $m$ weeks' attendance as [$N_{t-m}$, $N_{t-m+1}$, ..., $N_{t-1}$], and we call $m$ the \textit{memory window}.  Each strategy estimates the attendance at week $t$ as a linear combination of the previous $m$ weeks' attendance numbers, with the weights assigned to each week's attendance varying between strategies.  Thus, the $i$th strategy will estimate that attendance on week $t$ will be

\begin{equation}
    \hat{N}_{i,t} = \sum_{j=1}^m a_{i,j} N_{t-j} \,.
\end{equation} 

The weights assigned to each strategy are sampled from a distribution $f(\mathbf{a})$ where $\mathbf{a} = (a_1, a_2, \cdots, a_m)$.  In this paper, we take the $a_{i,j}$ to be independent and uniformly distributed in $[-1, 1]$.  The agents use a fixed cost function to evaluate the performance of each strategy.  The strategy with the lowest cost is selected, and the agent attends the bar at week $t$ if their chosen strategy predicts that attendance will be strictly less than the threshold.

Each agent must test their strategies on the outcomes of the $m$ prior weeks, and each predictor requires $m$ previous weeks of history.  Thus an initial seed history of length $2m$ is needed to run the model.  This seed history is generated by sampling each of the initial $2m$ attendance values independently from a uniform distribution over $\{0, 1, ..., A\}$.

\subsection{Meta-Strategies} \label{section:meta-strat}
One way to evaluate the cost of each strategy is to treat performance as a binary outcome.  Each strategy is satisfactory in week $t$ if the choice implied by its prediction (e.g., above or below the threshold) was the correct choice to make in that week.  The agent selects whichever strategy has the most correct choices in the previous $m$ weeks, and ties are broken at random.  We shall henceforth refer to this as the \textit{binary decision} meta-strategy.  Under this strategy, the cost at week $t$ for strategy $\textbf{a}$ is

\begin{equation}
\begin{aligned}
        C_t^\text{bin} (\mathbf{a}) &= \big| \big \{t-m \leq j < t \,\big| 
        \,(\hat{N}_{t-j} < T \leq N_{t-j}) \,  \\
        &\qquad\text{or}\;(\hat{N}_{t-j} \geq T > N_{t-j})
        \big\} \big| \,.
\end{aligned}
\label{Cost_Bin}
\end{equation}

The binary decision meta-strategy is used, for example in Refs.~\cite{chen2016, Chen2017, cara1999, RePEc:eee:phsmap:v:258:y:1998:i:1:p:230-236, stluce_2020}, while Refs.~\cite{ Whitehead2008, Challet2004} use a similar meta-strategy in which the cost is computed as an exponentially weighted mean over all time instead of equally weighting the costs of the $m$ previous weeks.

In contrast to the binary decision meta-strategy, agents could also select the strategy whose attendance predictions are closest to the actual recent history, regardless of whether they are above or below the threshold.  Thus, we consider a protocol in which agents select the strategy with the lowest squared error over the previous $m$ weeks, henceforth referred to as the \textit{error minimization} meta-strategy.  In this case, the cost for strategy $\mathbf{a}$ at week $t$ is given by

\begin{equation}
    \begin{aligned}
     C_t^\text{em}(\mathbf{a}) = (N_{t-1}& - \hat{N}_{t-1})^2 +(N_{t-2} - \hat{N}_{t-2})^2 + \\ 
     &... + (N_{t-m} - \hat{N}_{t-m})^2 \,.
     \end{aligned}
     \label{Cost_Errmin}
\end{equation}

The error minimization meta-strategy is less frequent in the literature, although various meta-strategies that rely on penalizing larger prediction errors have been used \cite{zambrano2004, cara1999, Fogel99}.

\section{Strategy Distribution and Steady State Behavior}

In this section we will study the dynamics of attendance under the two different meta-strategies we presented above.  An often-mentioned feature of the El Farol Bar Problem is that attendance fluctuates closely around the threshold $T$.  This is often taken as evidence of emergent self-organization.  However, as we shall see, situations where attendance oscillates wildly or settles above or below the threshold are also common depending on parameter choices.  We will illustrate these with numerical simulations and analytical calculations for both meta-strategies.

\subsection{\label{sec:bdms} Binary Decision Meta-Strategy}

\begin{figure*}[t]
    \centering
    \includegraphics[width=0.9 \textwidth]{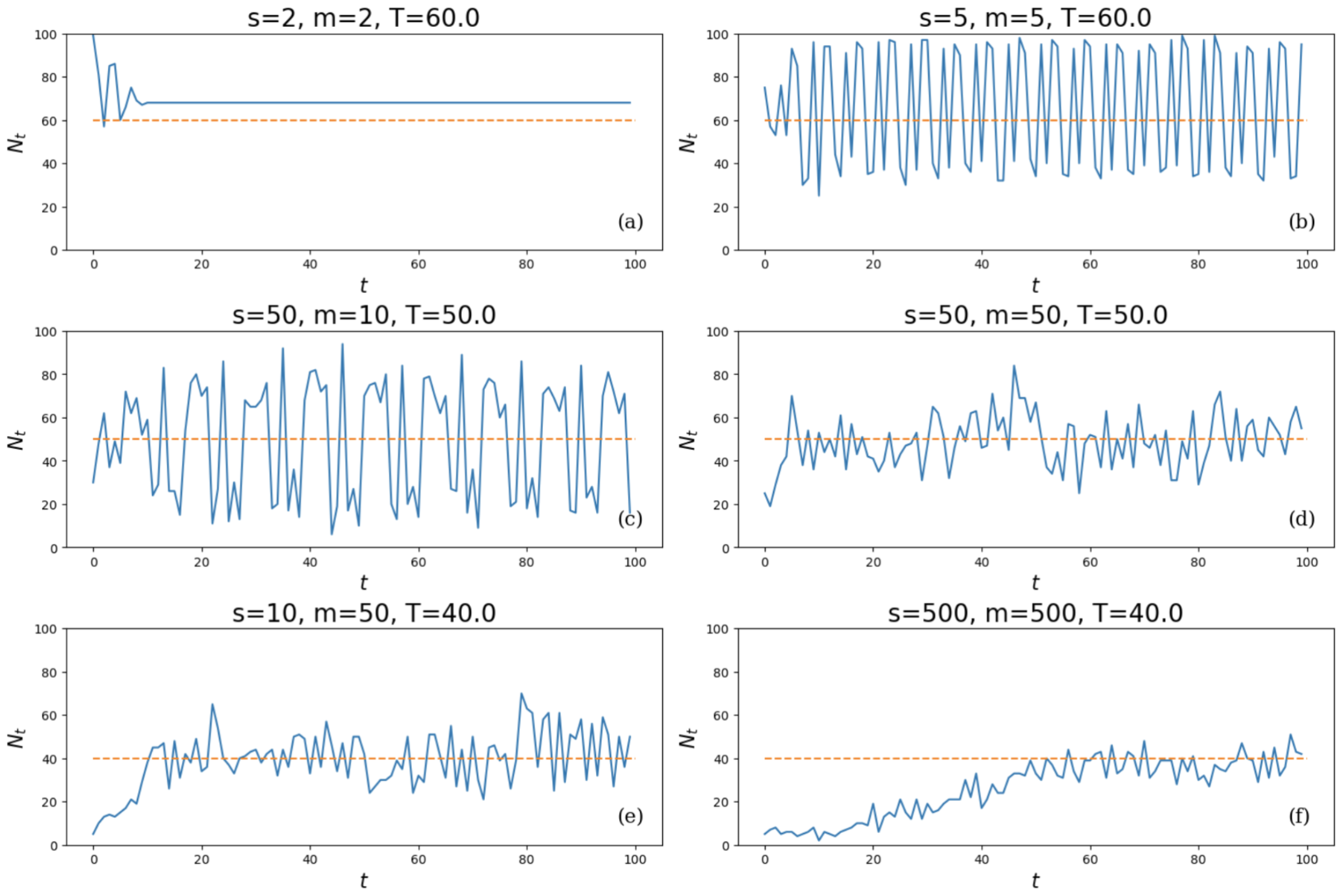}
    \caption{Attendance $N_t$ (solid blue curve) versus $t$ for the binary decision meta-strategy, with $A = 100$, and various values of $T$ (orange dashed line), $m$, and $s$.}
    \label{fig:choice_examples}
\end{figure*} 

Here we consider the binary decision meta-strategy introduced in Section \ref{section:meta-strat}.  Fig.~\ref{fig:choice_examples}  shows examples of attendance $N_t$ as a function of time for various combinations of memory window $m$ and strategy count $s$, and threshold $T$ (indicated with the dashed line), using $A = 100$.  For sufficiently low values of $m$ and $s$, attendance settles into a fixed point above the threshold, as shown in Fig.~\ref{fig:choice_examples}(a).  Considering that agents enjoy attending only when attendance is below the threshold, this is an unfavorable outcome for the system.  Figs.~\ref{fig:choice_examples}(b)-(f) show that for larger values of $m$ and $s$, attendance does fluctuate around the threshold, but fluctuations may be very large.

\begin{figure}[h]
    \centering
    \includegraphics[width=\columnwidth]{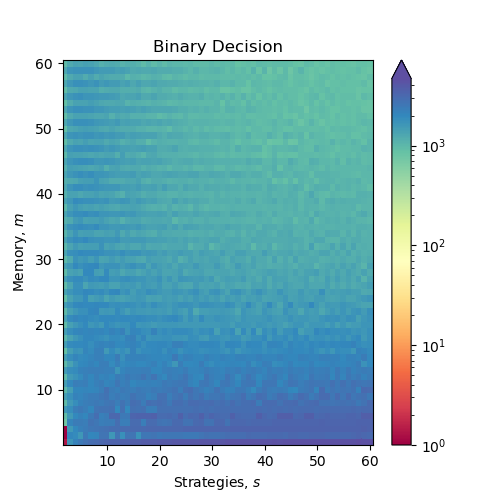}
    \caption{Standard deviation of attendance at steady state under the binary decision meta-strategy.  All simulations have $N = 10,000$, $T=6,000$.}
    \label{fig:grid_choice}
\end{figure}

Fig.~\ref{fig:grid_choice} shows broader trends in the relationship between $m$, $s$, and the size of fluctuations.  The color of each cell shows the standard deviation of attendance at steady state for the given values of $m$ and $s$.  Fixed points (i.e., $N_t$ remains constant for $t$ greater than some  $t_0$) are observed for some very small values of $m$ and $s$, visible as red cells in the lower left corner, while neighboring regimes have much larger fluctuations.  Large fluctuations are also an undesirable outcome since low attendance means resources (enjoyable bar seats) go to waste while high attendance means a large number of unsatisfied patrons.  As $m$ and $s$ grow, the size of fluctuations appears to decrease, as illustrated in Fig.~\ref{fig:choice_examples}(f).   This pattern is strikingly different from what is observed in some other variations of the problem, in which volatility increases as $s$ grows \cite{RePEc:eee:phsmap:v:258:y:1998:i:1:p:230-236, collins2017}.

First, we address the situation observed in Fig.~\ref{fig:choice_examples} (a), i.e., we determine for which parameters can attendance settle at a fixed point $N_t = N^*$.  We start by assuming that there is a fixed point $N^* \geq T$.  The case where $N^* < T$ is similar and will be discussed later.  Our strategy is to determine the probability, $P_\text{go}$, that an agent will attend the bar the next week.  Considering the limit $A \to \infty$, we then identify $P_\text{go}$ with $N^*/A$ to obtain an implicit equation for $N^*$.

For a given strategy $\mathbf{a}$, if $N_{t-j} = N^*$ for all $j \in \{1, 2, ..., m\}$,  the prediction for each time point in the memory window is simply

\begin{equation}
    \hat{N}_{i,t} = N^*\sum_{i=1}^m a_{ij} \,.
\end{equation}

According to the binary decision meta-strategy, if there is \emph{any}  strategy available to the agent such that $\hat{N}_{i,t} \geq T$, it will be selected.  So the agent will attend the bar only if all strategies have $\hat{N}_{i,t} < T$.  Thus the probability $P_\text{go}$ that an agent attends is

\begin{equation}
    \begin{aligned}
        P_\text{go} &= P(\hat{N}_{i,t} < T \; \forall\; i \in \{1,2, ..., s\})\\
        &= P(\hat{N}_{1,t} < T )^s \\
        &= P\left(\sum_{j=1}^m a_{1j} < \frac{T}{N^*} \right)^s \,.
    \end{aligned}
    \label{eq:pgo}
\end{equation}

We rewrite Eq.~\eqref{eq:pgo} in terms of an Irwin-Hall($m$) random variable, the sum of $m$ independent random variables, each uniformly distributed in $(0,1)$.  If $X$ has an Irwin-Hall($m$) distribution, $\sum_{j=1}^m a_{1j} \sim 2X - m$.  Hence,

\begin{equation*}
    \begin{aligned}
        P_\text{go} &= P\left(2X - m < \frac{T}{N^*} \right)^s \\
        &= P\left (X < \frac{T}{2N^*} + \frac{m}{2} \right )^s \\
        &= \left [ \frac{1}{m!} \sum_{k=0}^{\lfloor \frac{T}{2N^*} + \frac{m}{2} \rfloor} (-1)^k {m \choose k} \left (\frac{T}{2N^*} + \frac{m}{2} - k \right )^m  \right ]^s \,.
    \end{aligned}
\end{equation*}

Setting $P_\text{go} = N^*/A$ as discussed above, we obtain

\begin{equation}
    \begin{aligned}
    \frac{N^*}{A} = \bigg [ &\frac{1}{m!} \sum_{k=0}^{\lfloor \frac{T}{2N^*} + \frac{m}{2} \rfloor} (-1)^k {m \choose k} \\
    &\left (\frac{T}{2N^*} + \frac{m}{2} - k \right )^m  \bigg ] ^s,\qquad N^* \geq T\,.
    \end{aligned}
    \label{fp:above}
\end{equation}

\begin{figure}[h]
        \centering
        \includegraphics[width=\columnwidth]{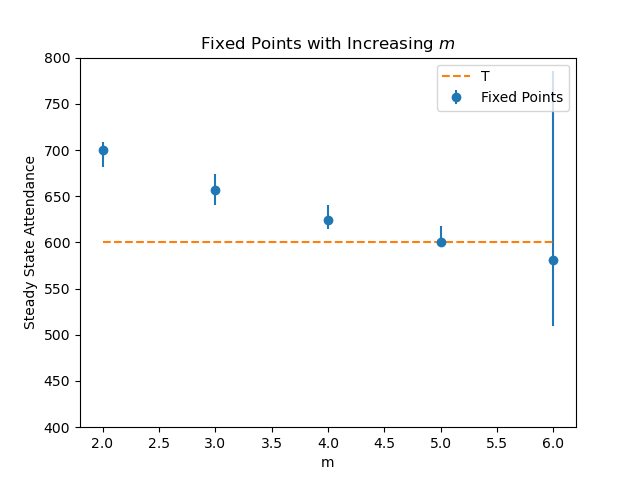}
        \hfill
        \includegraphics[width=\columnwidth]{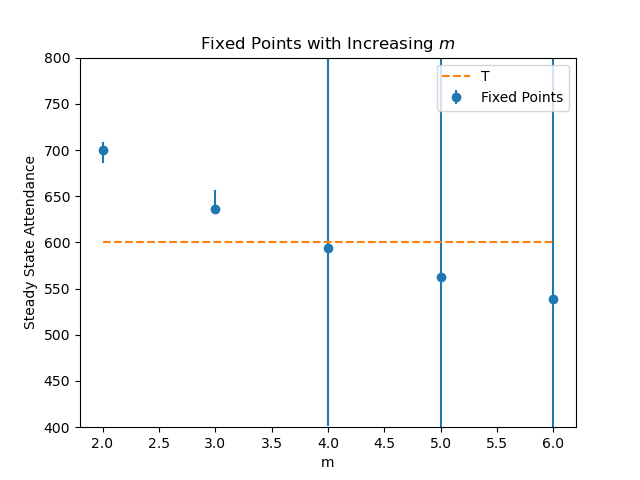}
        \begin{minipage}{.1cm}
        \vfill
        \end{minipage}
        \caption{Solutions to Eq.~\eqref{fp:above} (blue circles) as a function of $m$ for $s = 2$ (above) and as a function of $s$ for $m=2$ (below) with $A=1,000$ and $T=600$.  Error bars show the locations of fixed points in simulation when a fixed point is theoretically predicted (e.g. predicted fixed point is above $T$).  When fixed points are not predicted, error bars show the range of observed values in simulation.}
        \label{fig:fp_above}
\end{figure}

In Fig.~\ref{fig:fp_above}, the blue circles show the solutions to Eq.~\eqref{fp:above} as $m$ and $s$ increase.  The dashed line shows the threshold $T$.  For sufficiently low $m$ and $s$ values, solutions to Eq.~\eqref{fp:above} are above $T$, a necessary assumption to justify Eq.~\eqref{eq:pgo}.  Error bars on the plots in Fig.~\ref{fig:fp_above} show the minimum and maximum values over the last 50 time steps over 5 simulations of duration 100.  In the case of fixed points very close to $T$, such as $m=5$, $s=2$, some trials failed to converge and were discarded.  The error bar represents the spread of 5 trials for which attendance did converge above $T$.  As $s$ or $m$ grow sufficiently, the solution drops below $T$, and we no longer observe fixed points in simulation.  For these values, the error bars show the full spread of the observed oscillations over the last  50  trials.

The analysis for $N^* < T$ is similar (see Appendix \ref{app:bd_fixed_pts}), and it predicts there are no fixed points below the threshold, matching what we observe in simulation.

\begin{figure*}[ht]
    \centering
    \includegraphics[width=\textwidth]{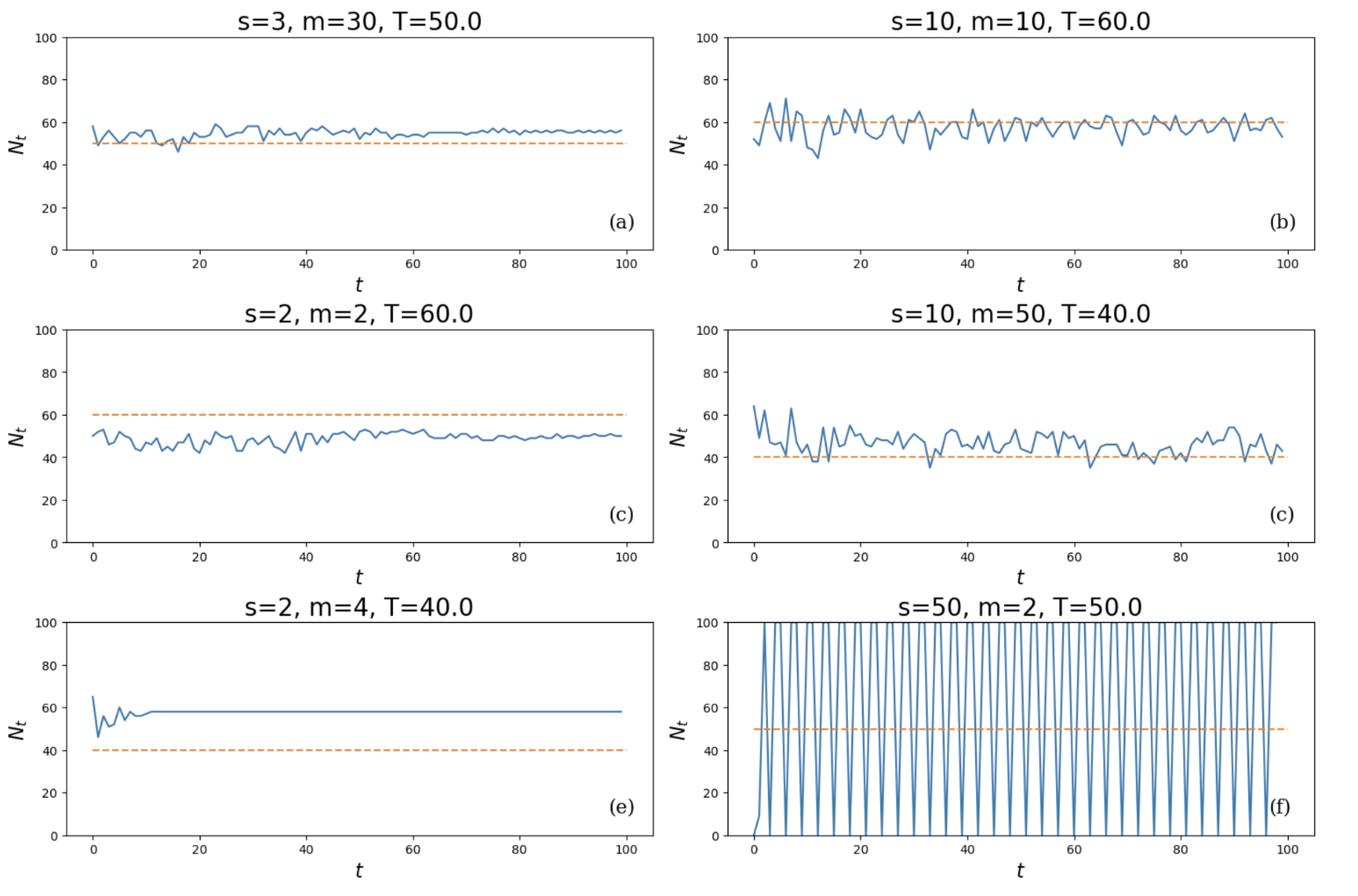}
    \caption{Attendance $N_t$ (solid blue curve) versus $t$ for the error minimization meta-strategy, with $A = 100$, and various values of $T$ (orange dashed line), $m$, and $s$.}
    \label{fig:errmin_examples}
\end{figure*}

Our analysis of the Binary Decision meta-strategy shows that, unless $m$ and $s$ are extremely large, attendance can either converge to a fixed point away from the threshold or undergo strong oscillations about it.  As far as we know, fixed points for the Binary Decision meta-strategy have not been studied previously.  For large $m$ and $s$, our numerical results suggest that attendance does indeed converge to the threshold $T$.  Overall, our results for the binary decision meta-strategy highlight the important effects of the number of strategies $s$ and the memory window $m$ on the collective dynamics of the agents.  

\subsection{Error Minimization Meta-Strategy}

\begin{figure}
    \centering
    \includegraphics[width=\columnwidth]{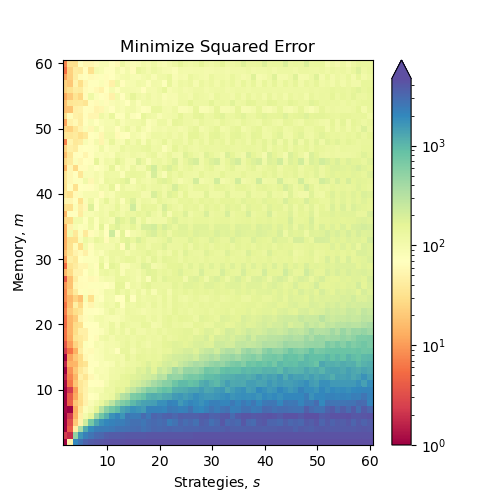}
    \caption{Standard deviation of attendance at steady state under the minimize squared errors meta-strategy.  All simulations have $N = 10,000$, $T=6,000$.}  
    \label{fig:grid}
\end{figure}

Now we study the error minimization meta-strategy.  In Fig.~\ref{fig:errmin_examples} we show attendance  $N_t$ versus time $t$ for different values of $m$, $s$, and $T$.   As in the binary decision meta-strategy, there are some cases where attendance settles at an approximately constant  value.  When it does so fluctuations are, in general, smaller than in the binary decision case.  On the other hand, when the number of strategies $s$ sufficiently exceeds the memory window $m$, as it does in Fig.~\ref{fig:errmin_examples}(f), attendance can have very large oscillations, in some cases alternating between $0$ and $A$.  This is illustrated in Fig.~\ref{fig:grid}, which shows the standard deviation of steady state attendance as a function of $m$ and $s$.  Volatility appears to increase as $s$ grows relative to a fixed $m$, consistent with patterns observed by Johnson et. al and Collins \cite{RePEc:eee:phsmap:v:258:y:1998:i:1:p:230-236, collins2017}.  In the remainder of this section, we investigate the following two questions:  (i) why does attendance sometimes settle around values other  than the threshold $T$?, and (ii) why does the availability of more strategies for each agent (larger $s$) lead  to larger oscillations in attendance?

To address question (i), we solve for fixed points in weekly attendance by assuming a fixed point, $N^*$, exists and determining the probability, $P_\text{go}$ that each agent would attend the bar.  As we did with the binary decision meta-strategy, we consider the limit $A \to \infty$ and identify $P_\text{go}$ with $N^*/A$ to derive an implicit equation for $N^*$.  Supposing attendance every week is $N^*$, that is $N_t = N^*$ for all $t$, the cost function for each week simplifies to

\begin{equation}
    C_t(\mathbf{a}) = m (N^*)^2 \left(\sum_{j=1}^m a_{i,j} - 1\right)^2 \,.
\end{equation}

Hence, the optimal strategy is simply the one that minimizes the quadratic form $(\sum_{j=1}^m a_{i,j} - 1)^2$.  In the following, we denote by $\mathbf{a}^*$ the strategy weights selected by an arbitrary agent among their $s$ available choices, and by $f^*$ their probability density function.  In Appendix \ref{app:dist_err}, we show that this probability density is

\begin{equation}
    f^*(\mathbf{a}^*) = \frac{s}{2^m}\left(1 - \int_{E(\mathbf{a}^*)} \frac{1}{2^{m}} d\mathbf{a}\right)^{s-1} \,,
    \label{pdf_mse}
\end{equation}

where $E(\mathbf{a}^*)$ is the region of strategy space that would have a lower cost than $\mathbf{a}^*$, i.e.,

\begin{equation}
    \begin{aligned}
        E(\mathbf{a}^*) = \{\mathbf{a}: C(\mathbf{a}) < C(\mathbf{a}^*)\} .
    \end{aligned}
    \label{ellipse}
\end{equation}

In order to simplify the calculations while still gaining insight into the dynamics of the system, we study the case where $m = 2$  in detail.  In this case, the PDF of the weights, given by Eq.~\eqref{pdf_mse}, evaluates to

\begin{equation}
    f^{*}(\mathbf{a}^*) = \frac{s}{4}\left[ 1 - \frac{1}{4} g(\mathbf{a}^*)\right]^{s-1}  \,,
    \label{eq:fpdensity}
\end{equation}

where

\begin{equation}
    g(\mathbf{a}^*) = \begin{cases}
        2|a_1^* + a_2^* - 1| & \text{if} \ |a_1^* + a_2^* - 1| < 1, \\
        \frac{1}{2}[6|a_1 ^* + a_2^* - 1| - &\\
        (a_1^* + a_2^* - 1)^2 - 1] & \text{otherwise}.
    \end{cases}
\end{equation}

At the fixed point, an agent will attend the bar if their chosen strategy, $\mathbf{a^*}$, satisfies 

\begin{equation}
    \begin{aligned}
        N^* \sum_{j=1}^m a_j^* < T \,.
    \end{aligned}
    \label{eq:condition}
\end{equation}

Hence, we compute $P_\text{go}$ by integrating Eq.~\eqref{eq:fpdensity} over the region satisfying Eq.~\eqref{eq:condition}.  With $s=2$ and $m = 2$, as assumed in Eq.~\eqref{eq:fpdensity}, this evaluates to

 \begin{equation}
     P_\text{go} = \begin{cases}
     \frac{1}{24} \left [ 6 + 12 \frac{T}{N^*} + 3 \frac{T}{N^*}^2 - 2\frac{T}{N^*}^3 \right ] & 0 < \frac{T}{N^*} \leq 1 \\[5pt]
     \frac{1}{24} \left [ -4 + 36 \frac{T}{N^*} -15 \frac{T}{N^*}^2 + 2\frac{T}{N^*}^3 \right ] & \text{otherwise}.
    \end{cases}
    \label{eq:s2m2}
 \end{equation}

We are now prepared to construct an implicit equation for $N^*$ by setting $P_\text{go} = N^* / A$:

    \begin{figure}
        \centering
        \includegraphics[width=\columnwidth]{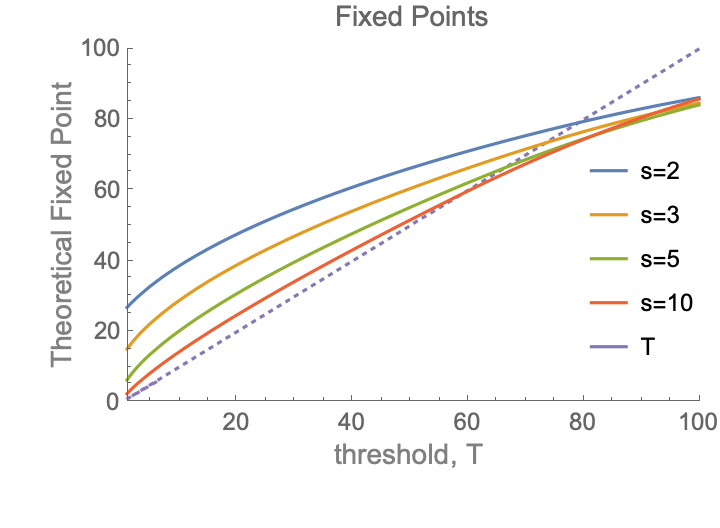}
        \caption{Fixed points predicted by numerical solution of Eqs.~\eqref{eq:s2m2}, \eqref{eq:fps3m2}, \eqref{eq:fps5m2}, \eqref{eq:fps10m2}, as a function of $T$ for $A=100$, $m=2$ and $s=2,3,5,10$.  The identity is shown as a dashed line.}
        \label{fig:fp_vs_T}
    \end{figure}
    
   \begin{figure}[t]
        \centering
        \includegraphics[width=\columnwidth]{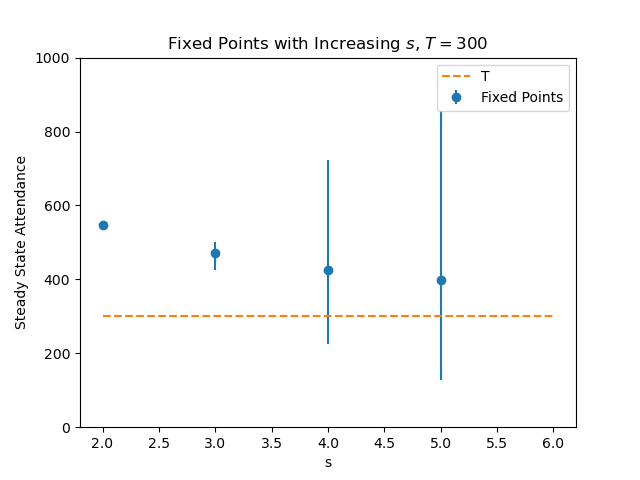}
        \hfill
        \includegraphics[width=\columnwidth]{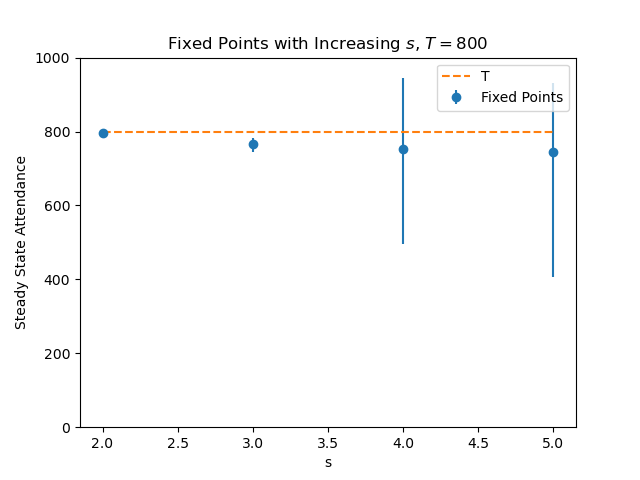}
        \begin{minipage}{.1cm}
        \vfill
        \end{minipage}
        \caption{Solutions to Eq.~\eqref{eq:fpm2} (blue circles) as a function of $s$ for $m = 2$, $A=1,000$, and $T = 300$ (above) and $T = 800$ (below).  Error bars show the observed range of values $N_t$ in simulation over time steps $t = 50$ through $t = 100$ in five independent simulations.}
        \label{fig:fp_locs}
\end{figure}

 \begin{equation}
     \frac{N^*}{A} = \begin{cases}
     \frac{1}{24} \left [ 6 + 12 \frac{T}{N^*} + 3 \frac{T}{N^*}^2 - 2\frac{T}{N^*}^3 \right ] & 0 < \frac{T}{N^*} \leq 1 \\[5pt]
     \frac{1}{24} \left [ -4 + 36 \frac{T}{N^*} -15 \frac{T}{N^*}^2 + 2\frac{T}{N^*}^3 \right ] & \text{otherwise}.
     \end{cases}
    \label{eq:fpm2}
 \end{equation}
 
Eq.~\eqref{eq:fpm2} is an implicit equation for the fixed point $N^*$ as a function of the threshold $T$, for $m =2$, $s=2$.  While the calculations become more cumbersome as $s$ is increased, we have obtained analogous expressions for $m = 2$  and $s = 3, 4, 5$ (see Appendix \ref{app:fp_errmin}).  Plotting the fixed point $N^*$ as a function of $T$ in Fig.~\ref{fig:fp_vs_T}, we note that the fixed point can be either above or below the threshold, depending on other parameters.  Therefore, for this particular meta-strategy, attendance does  \emph{not} self-organize around the threshold, as is often claimed for El Farol-type problems.  In Refs. \cite{LUS2005651, RePEc:eee:phsmap:v:258:y:1998:i:1:p:230-236} it was observed numerically that in certain versions of El Farol Bar Problem the mean attendance can differ from the attendance threshold. Even though our implementation of the El Farol Bar Problem is different, our model and results provide an example where this issue can be explored theoretically.

While we obtain implicit expressions for fixed points, $N^*$, our analysis does not give any information on their stability.  In some cases, instead of the fixed points predicted from our analysis, we observe large fluctuations in attendance, and we hypothesize that this is  because the fixed points are unstable.  This is illustrated in Fig.~\ref{fig:fp_locs}, which shows our theoretical predictions for the fixed points (blue circles), the threshold $T$ (dashed line), and the range over which attendance varies over the last 50 time steps of 5 different simulations of length 100 (error bars).  When this range is small, suggesting a fixed point, the numerical  values of $N_t$ agree well with our theoretical result.  For larger values of $s$, the fixed point appears to lose stability and large fluctuations are observed.  

 We now address question (ii), namely, why do large oscillations occur as $s$ becomes large.  Recall that the probability density for strategies selected by an agent is given by Eq.~\eqref{eq:fpdensity}.  We observe that the area of $E(\mathbf{a}^*)$ and thus the value of $\int_{E(\mathbf{a}^*)}\frac{1}{2^{m}} d\mathbf{a}$ will equal 0 only when the strategy $\mathbf{a}^*$ minimizes the cost function.  This minimizer is unique since the cost function is quadratic.  For all other choices of $\mathbf{a}^*$, this integral will be strictly between 0 and 1.  Hence, as $s$ approaches $\infty$, $f^*$ approaches the Dirac delta function centered at the optimal strategy.  Thus, we expect that as $s$ approaches infinity all agents will make the same choice and every iteration will have either all agents or no agents attending.  Interestingly, improved individual predictive accuracy leads to decreased global utility.

\begin{figure}[t]
    \includegraphics[width=0.5\textwidth]{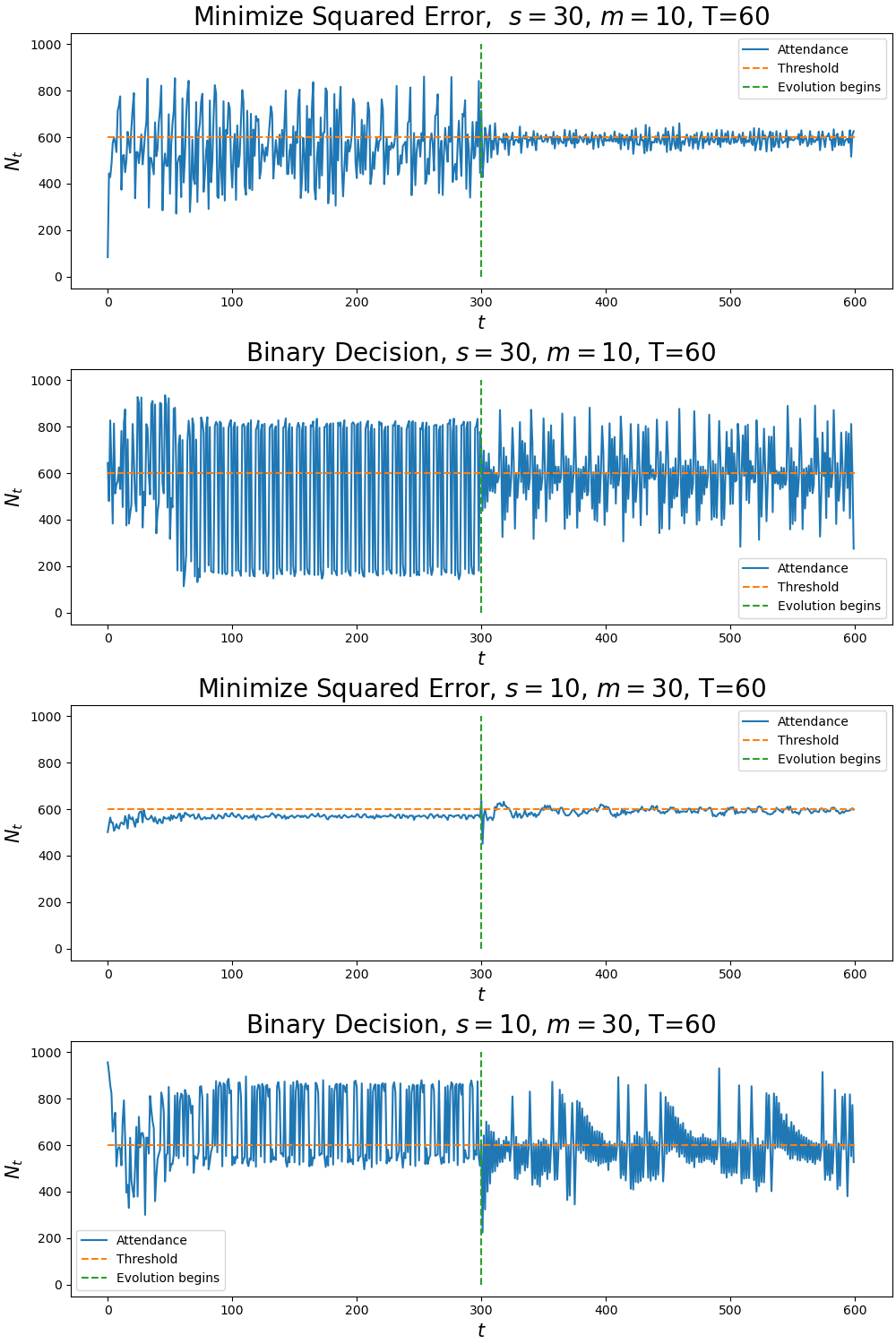}
    \caption{The effect of evolution in different parameter ranges.  Evolution begins at $t = 300$.  All simulations were run with $\lambda = 0.2$, $\epsilon = 0.7$, $A =  1,000$.}
    \label{fig:evolution_on_off}
\end{figure}

\section{Evolution}

\begin{figure}[t]
    \includegraphics[width=\columnwidth]{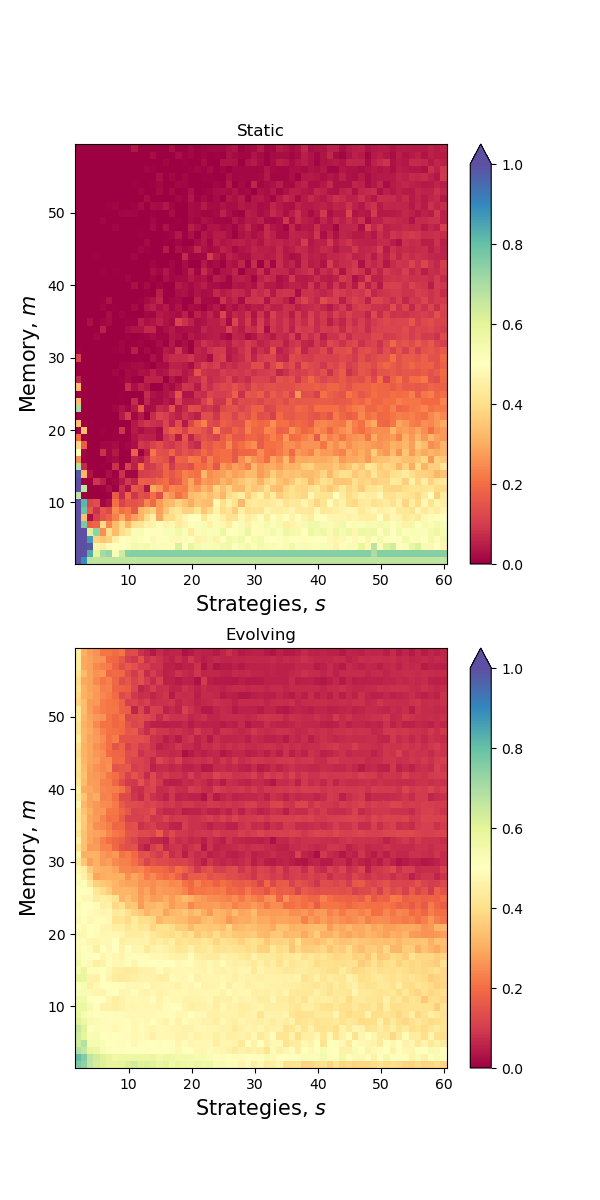}
    \caption{Fraction of time above $T$ at steady state with the error minimization meta-strategy, with $T = 6,000$, $A = 10,000$, $\epsilon = 0.5$, $\lambda = 0.2$, without (top) and with (bottom) evolution.}
    \label{fig:time_above}
\end{figure}

\begin{figure}
    \includegraphics[width=\columnwidth]{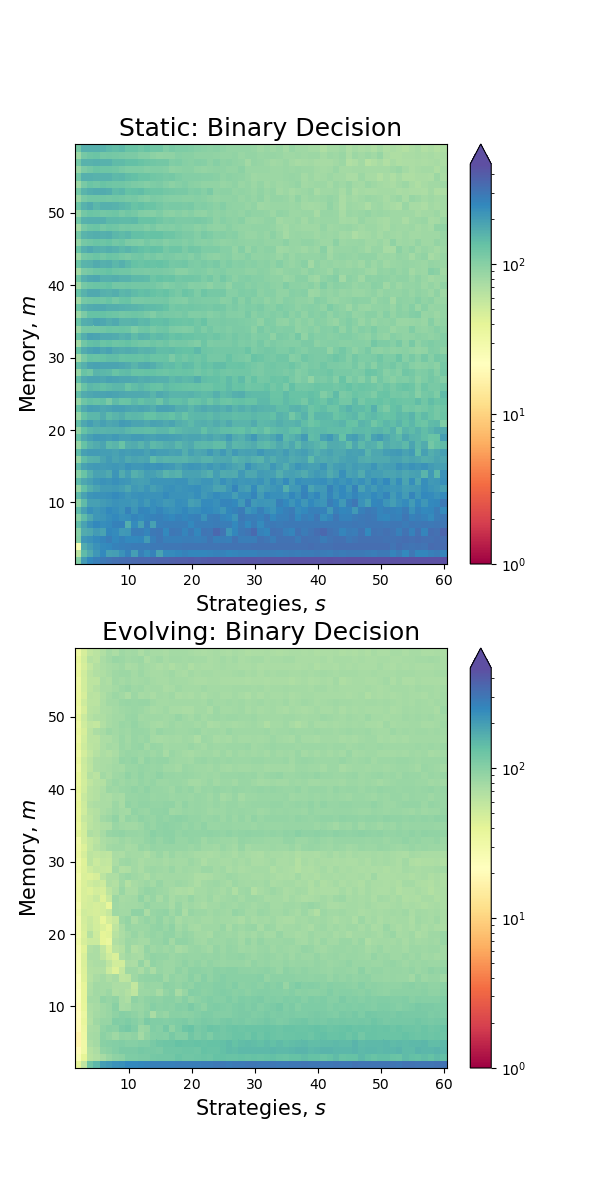}
    \caption{Average distance from the threshold at steady state with the binary decision meta-strategy, without (top) and with (bottom) evolution.}
    \label{fig:grid_evolving_choice}
\end{figure}

Using the binary decision meta-strategy, attendance fluctuates around the threshold in all but a few anomalous regimes with very few strategies, but the size of fluctuations remains large unless both $m$ and $s$ are extremely large.  Meanwhile, the error minimization meta-strategy only oscillates close to the threshold when $s$ is in an optimal range relative to $m$; too low, and the system gets caught in a fixed point away from the threshold, or too high, and fluctuations become excessively large, leading to poor utilization of the resource.

Different evolutionary paradigms have been proposed, leading to dramatically varied results.  When agents use stochastic strategies and are able to adjust their probability of attendance, the system can converge to a Nash equilibrium in which all agents attend with probability $T/A$ \cite{Miklos2012}.  However, when agents are able to adjust their own parameters in a deterministic strategy space, they may end up in a deadlock where some agents always attend and other agents almost never attend \cite{Whitehead2008}, or, alternatively, it may drive increased volatility \cite{Fogel99}, just as we observe when agents are given access to an increased number of strategies and agents' decisions thus synchronize.

Instead, we propose a different approach to evolution in the agent-based system; rather than refining individual strategies, agents may re-draw their own $m$ and $s$ values if they are not getting sufficient utility in the present regime.  Every time they change $m$ and $s$, the weights for each strategy are drawn once again from a uniform distribution in $[-1,1]$.  The strategies then remain fixed until the next time the agent redraws $m$ and $s$.

The agent gives each strategy an initial trial period of length $2m$, whatever their current $m$ value is.  After that, on each step, the agent computes their performance over a window of length $m$.  If their fraction of correct choices is less than $\epsilon$, the agents redraw $m$ and $s$ with probability $\lambda$.  In the simulations that follow, we give each agent the same starting $m_0$ and $s_0$ values, and allow the agents to draw uniformly at random from the range $s \in [s_0/2, 3s_0/2]$ and $m \in [m_0/2, 3m_0/2]$.  Fig.~\ref{fig:evolution_on_off} shows the effect of evolution in different parameter regimes, for $\lambda = 0.2$, $\epsilon = 0.7$ (as we will discuss below, choices of $\lambda$ and $\epsilon$ do not significantly affect the results).  In all panels, the simulation is run without evolution for the first 300 steps.  Then evolution is turned on a $t=300$ (vertical dashed line). The solid blue line shows the attendance, $N_t$, and the horizontal dashed line shows the threshold, $T$.  

Evolution appears to both reduce the amplitude of large fluctuations and force attendance to cross the threshold when it otherwise would not.  If attendance were to remain above $T$ for a sufficiently long period of time relative to the agents' $m$ values, more than $T$ agents would be dissatisfied with their current regime, and thus they would be likely redraw their parameters until the behavior of the systems switches.  Likewise, if attendance remains below $T$ for sufficiently long, then $A - T$ agents would be dissatisfied at each time step, thus redrawing their parameters.  Evolution also appears to limit the size of fluctuations.  In a scenario with large fluctuations, many agents will be consistently dissatisfied with their outcomes, as they will be attending in over-crowded weeks and absent in weeks with capacity.  Thus, scenarios with high fluctuations will be unstable.  

Fig.~\ref{fig:time_above} shows the fraction of time attendance is above $T$ at steady state with the error minimization meta-strategy for a range of parameters without (above) and with (below) evolution.  When an evolutionary paradigm is used, a larger region of strategy space has attendance close above  $T$ around half the time.  Even in regions where attendance rarely crosses $T$, the fraction never reaches 0, while without evolution there are large regions of strategy space for which attendance never rises above the threshold.  With the binary decision meta-strategy, the fraction of time above the threshold remains close to  $1/2$, with or without evolution (not shown).

\begin{figure}
    \includegraphics[width=\columnwidth]{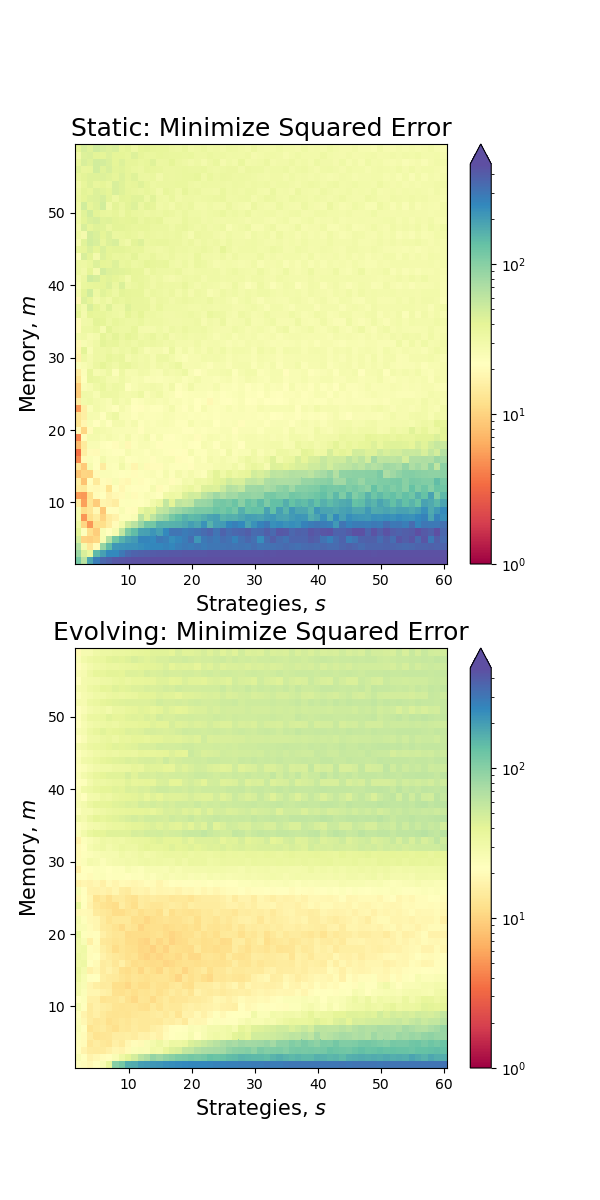}
    \caption{Average distance from the threshold at steady state with the error minimization meta-strategy, without (top) and with (bottom) evolution.}
    \label{fig:grid_evolving_mse}
\end{figure}

Figs. \ref{fig:grid_evolving_choice} and \ref{fig:grid_evolving_mse} show the effect of evolution on the average distance from the threshold, $T$, at steady state, calculated as the time average of $|N_t - T|$.  Evolution brings attendance closer to $T$, particularly in regions with larger fluctuations.  Note that while these plots look similar to Figs. \ref{fig:grid_choice} and \ref{fig:grid}, distance from threshold is a different metric than standard deviation of attendance.

\begin{figure}
    \centering
    \includegraphics[width=\columnwidth]{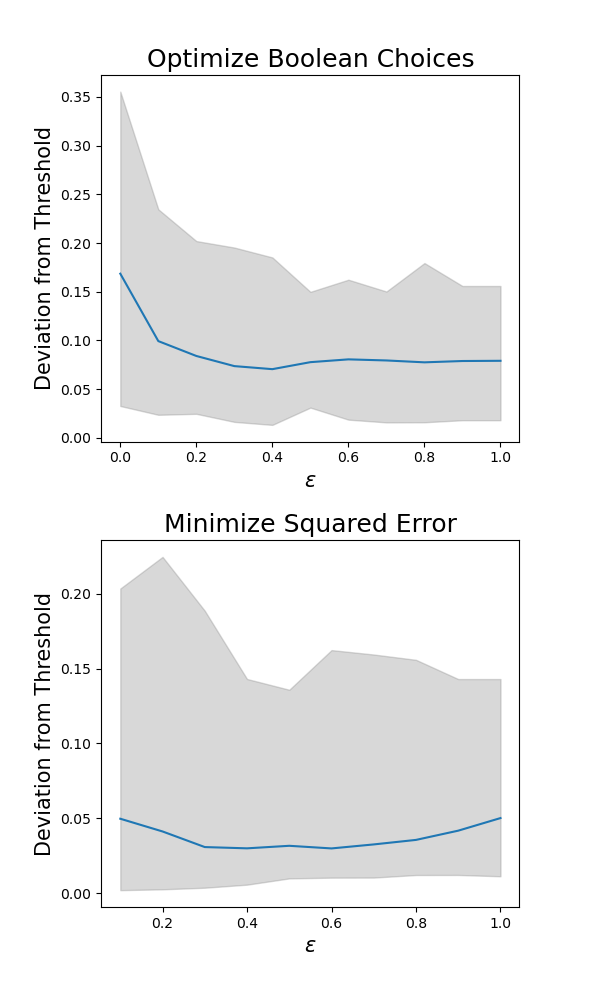}
    \caption{Mean, minimum, and maximum distance from threshold, $T$, at steady state for different values of $\epsilon$.  Five trials each were run on $m$ and $s$ values of 5, 10, 20, and 50.}
    \label{fig:epsilon}
\end{figure}

Our evolutionary model does introduce two new parameters, $\lambda$ and $\epsilon$, but neither appears to have a dramatic effect on the behavior of the system.  Fig.~\ref{fig:epsilon} summarizes the deviation from the threshold at different values of $\epsilon$ for a range of $m$ and $s$ values.  The effects are not dramatic for $\epsilon \geq 0.2$.  The effect of $\lambda$ is even less pronounced for $\lambda > 0$, but including this parameter rather than setting it to 1 allows for a smoother beginning to the simulation as all agents do not simultaneously change parameters.

\section{Discussion}

One of the most intriguing features that we observe in both non-evolving regimes of the El Farol Bar Problem is that systems in which individual agents' models appear to make better predictions based on past history end up having worse utilization of the resource in question.  We have seen mechanistically how increasing the number of strategies available to an agent, given a fixed memory length, drives volatility.  

From a more philosophical perspective, we have an interesting case study on the choice of metrics to optimize.  In Arthur's original paper, the El Farol Bar Problem is presented as an example of how a collection of agents making rational decisions based on incomplete information and flawed models leads to an adaptive self-organization emerging in the system \cite{Arthur94}.  The mechanism underlying this self-organization is that any strategy that consistently performs poorly will not be selected.  However, our and other papers have shown that giving agents access to strategies that appear to perform better based on past performance does not lead to a better outcome for the system as a whole \cite{Challet2004, Fogel99, collins2017}.  When agents are able to all adjust their strategies in almost identical ways, the individual strategies selected change, but volatility in the system remains consistent.  Thus, the adaptive self-organization observed by Arthur is dependent on constraints in the agents' access to predictive models that are, by their own standards, desirable.

Other works that have included evolutionary algorithms have used that evolution to improve prediction accuracy, leading to sub-optimal outcomes for the common sharing of the resource \cite{Fogel99, Whitehead2008}.  However, in our approach to evolution, we attempt to optimize directly for agents' ability to utilize the resource.

The effect of the evolutionary paradigm is not to drive agents toward more accurate predictions, but to drive the system toward more heterogeneity.  This heterogeneity, rather than agent-level prediction accuracy, creates a system in which agents experience higher rates of utility in a more stable and equitable way.

\appendix

\section{\label{app:bd_fixed_pts} Fixed Points Below $T$ with the Binary Decision Meta-Strategy}

This  analysis follows similar logic to the case where the fixed point is above $T$, discussed in Section \ref{sec:bdms}.  If attendance is below the threshold, an agent will attend if they have \emph{at least} one strategy that predicts $\hat{N}_{i,t} < T$.  So in this case, we have

\begin{equation}
    \begin{aligned}
        P_\text{go} &= P(\exists \;i  \in \{1,2, ..., s\} \,,\, \hat{N}_{i,t} < T)\\
        &= 1 - P(\hat{N}_{i,t} \geq T \;\forall \; i  \in \{1,2, ..., s\}) \\
        &= 1 - P\left(\sum_{j=1}^m a_{1j} \geq \frac{T}{N^*} \right)^s \,.
    \end{aligned}
\end{equation}

which, after evaluating the probability and using the self-consistent condition $P_\text{go} = N^*/A$, yields the implicit equation

\begin{equation}
    \begin{aligned}
    \frac{N^*}{A} = 1 - \bigg [&1 - \frac{1}{m!} \sum_{k=0}^{\lfloor \frac{T}{2N^*} + \frac{m}{2} \rfloor} (-1)^k {m \choose k} \\
    &\left (\frac{T}{2N^*} + \frac{m}{2} - k \right )^m  \bigg]^s,\qquad N^* < T\,.
    \end{aligned}
    \label{fp:below}
\end{equation}

\begin{figure}[h]
        \centering
        \includegraphics[width=\columnwidth]{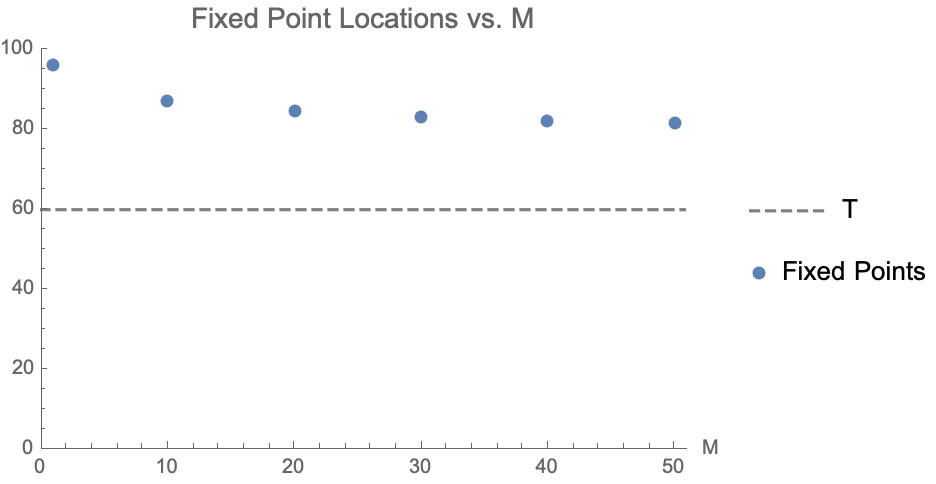}
        \hfill
        \includegraphics[width=\columnwidth]{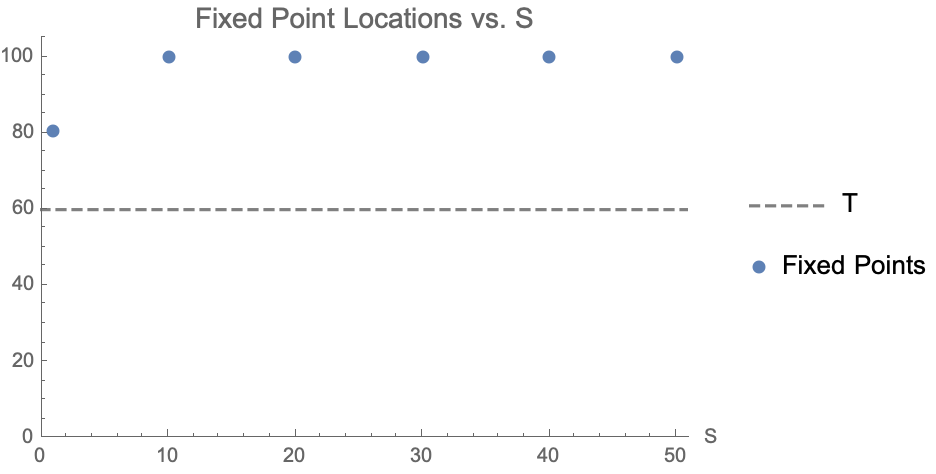}
        \begin{minipage}{.1cm}
        \vfill
        \end{minipage}
        \caption{Solutions to equation \eqref{fp:below} as a function of $m$ (above) and $s$ (below).}
        \label{fig:fp_below}
\end{figure}

Fig.~\ref{fig:fp_below} shows solutions to Eq.~\eqref{fp:below} as $m$ and $s$ increase.  Solutions to Eq.~\eqref{fp:below} were above $T$ with all parameters tested, and hence inconsistent with the setup.  No fixed points were observed below  $T$ in simulation either.  While it is difficult to interpret this necessity from equation \ref{fp:below}, we observe that with the uniform $[-1, 1]$ distribution of weights, strategies are biased toward underestimating attendance.  In order to have a fixed point above below $T$, at least $\frac{N - T}{N}$ agents would need to overestimate attendance.

\section{\label{app:dist_err} Distribution of Strategies Chosen in Error Minimization Meta-Strategy}

We consider the distribution of strategies selected by each agent under the minimize squared prediction error.  The cost function for each strategy is given by Eq.~\eqref{Cost_Errmin}.  The cost function is quadratic and will always have a unique minimizer unless all $N$'s are equal to 0. Depending on the attendance history, this minimum could be inside or outside of the base of support of the $a$ values.  

However, there must also be a unique minimizer of Eq.~\eqref{Cost_Errmin} within the base of support of the $a$'s. If the optimal strategy is within the range of possible $a$ values, this will be the unique minimizer.  We now consider the case where the the minimizer of \eqref{Cost_Errmin} is outside of the base of support. Taking $c_\text{min}$ to be the lowest cost observed within the base of support of the $a$'s, consider $\ell_\text{min}$ the level curve of the cost function at $c_\text{min}$.  Level curves of equation \eqref{Cost_Errmin} are all elliptical, while the base of support of the $a$ values is rectangular.  If $\ell_\text{min}$ intersected the base of support of the $a$'s at multiple points, there would be some region between $\ell_\text{min}$ and the true minimizer of the cost function within the base of support.  Since the cost surface is convex, this region would have cost below $C_\text{min}$, contradicting our construction of $C_\text{min}$.

Consider the cumulative distribution of strategies selected.  We will use $F(\mathbf{a})$ to denote the cumulative distribution of strategies randomly assigned to  an agent, $F^*(\mathbf{a})$ to denote the cumulative distribution function of the strategy selected, and $a_{k*}$ to denote the $k$th weight of the selected strategy.  Since the strategies are i.i.d.,

\begin{equation}
    \begin{aligned}
F^*(\mathbf{a}) &= P(a_{1*} \leq a_1, \cdots, a_{m*} \leq a_m)\\
&= s P\big(a_{1,1} < a_1,\cdots, a_{m,1} < a_m, \\ 
&\qquad C(\mathbf{a}_1) < \min\{C(\mathbf{a}_j) | j \in 2, ..., s \}\big) \,.
    \end{aligned}
\end{equation}

Conditioning on the values of the optimal strategy chosen, we find that 

\begin{widetext}
\begin{equation}
    \begin{aligned}
        F^*(\mathbf{a}) &= s \int_{-\infty}^{a_1} \cdots \int_{-\infty}^{a_m} P(C(\mathbf{a}_1) < \min\{  C(\mathbf{a}_j) | j \in 2,\cdots, s \} \,|\, \mathbf{a} = \mathbf{b}) f(\mathbf{b}) d\mathbf{b}\\
        &=  s \int_{-\infty}^{a_1} \cdots \int_{-\infty}^{a_m} \left(1 -  \int_{E(\mathbf{b})} f(\mathbf{c})  d\mathbf{c}\right)^{s - 1} f(\mathbf{b}) d \mathbf{b} \,.
    \end{aligned}
\end{equation}
\end{widetext}

Thus, when strategies are sampled from the Uniform(-1,1) distribution, the PDF of the chosen strategies, ($a_1^*$, $a_2^*$, $\cdots$, $a_m^*$) is

\begin{equation}
    f^*(\mathbf{a}) = \frac{s}{2^m}\left(1 - \int_{E(\mathbf{a})} \frac{1}{2^{m}} d\mathbf{a}\right)^{s-1} \,.
\end{equation}

\section{\label{app:fp_errmin} Implicit Equations for Fixed Points with the Error Minimization Meta-Strategy}

 Here, we show the implicit equations for additional fixed points shown in Fig.~\ref{fig:fp_vs_T}.  For each choice of $s$, we compute $P_\text{go}$ by integrating Eq.~\eqref{eq:fpdensity} over the region satisfying Eq.~\eqref{eq:condition}, and derive the implicit equation by setting $P_\text{go} = N^* / A$.  With $s=3$ and $m = 2$, we have

 \begin{equation}
    \frac{N^*}{A} = \begin{cases}
         \frac{1}{64} \big[-3 \left(\frac{T}{N^*}\right)^4+18 \left(\frac{T}{N^*}\right)^2+ & 0<\frac{T}{N^*}\leq 1\,, \\
         \qquad24 \frac{T}{N^*}+8\big] & \\
         \frac{1}{64} \big[-3 \left(\frac{T}{N^*}\right)^4+32 \left(\frac{T}{N^*}\right)^3- & \text{otherwise} \,. \\
         \qquad 126 \left(\frac{T}{N^*}\right)^2 + 216 \frac{T}{N^*}-72\big] &  
    \end{cases}
    \label{eq:fps3m2}
 \end{equation}

 With $s = 5$, $m = 2$, we have

 \begin{equation}
     \frac{N^*}{A} = \begin{cases}
         \frac{1}{384} \big[-5 \left(\frac{T}{N^*}\right)^6-12 \left(\frac{T}{N^*}\right)^5 + & 0<\frac{T}{N^*}\leq 1 \,,\\
         \qquad15 \left(\frac{T}{N^*}\right)^4+80 \left(\frac{T}{N^*}\right)^3 & \\
         \qquad + 105 \left(\frac{T}{N^*}\right)^2+60 \frac{T}{N^*}+12\big] &\\
         \frac{1}{384} \big[-5 \left(\frac{T}{N^*}\right)^6+84 \left(\frac{T}{N^*}\right)^5-&\\
         \qquad585 \left(\frac{T}{N^*}\right)^4+2160 \left(\frac{T}{N^*}\right)^3-&\\
         \qquad 4455 \left(\frac{T}{N^*}\right)^2+4860 \frac{T}{N^*}-1804\big]
           & \text{otherwise} \,.
        \end{cases}
        \label{eq:fps5m2}
 \end{equation}

 Finally, with $s=2$, $m=10$, we have

 \begin{equation}
    \frac{N^*}{A} = \begin{cases}
     \frac{1}{{22528}}\big[-10 \left(\frac{T}{N^*}\right)^{11}-& 0<\frac{T}{N^*}\leq 1\\
     \qquad 77 \left(\frac{T}{N^*}\right)^{10}-220 \left(\frac{T}{N^*}\right)^9 -&\\
     \qquad165 \left(\frac{T}{N^*}\right)^8+660 \left(\frac{T}{N^*}\right)^7+&\\
     \qquad 2,310 \left(\frac{T}{N^*}\right)^6+3,696 \left(\frac{T}{N^*}\right)^5+&\\
     \qquad3,630\left(\frac{T}{N^*}\right)^4+2,310 \left(\frac{T}{N^*}\right)^3+&\\
     \qquad935 \left(\frac{T}{N^*}\right)^2+220 \frac{T}{N^*}+22\big] &  \\
     \frac{1}{22528}\big[10 \left(\frac{T}{N^*}\right)^{11}-&\text{otherwise}\,.\\
     \qquad 319 \left(\frac{T}{N^*}\right)^{10}+4,620 \left(\frac{T}{N^*}\right)^9-&\\
     \qquad40,095 \left(\frac{T}{N^*}\right)^8+&\\
     \qquad231,660 \left(\frac{T}{N^*}\right)^7-&\\
     \qquad935,550 \left(\frac{T}{N^*}\right)^6+&\\
     \qquad2,694,384\left(\frac{T}{N^*}\right)^5-&\\
     \qquad5,533,110 \left(\frac{T}{N^*}\right)^4+&\\
     \qquad7,938,810 \left(\frac{T}{N^*}\right)^3-&\\
     \qquad7,577,955 \left(\frac{T}{N^*}\right)^2+&\\
     \qquad4,330,260 \frac{T}{N^*}-1,099,404\big] &
    \end{cases}
     \label{eq:fps10m2}
 \end{equation}

\bibliographystyle{plain} % We choose the "plain" reference style
\bibliography{refs} % Entries are in the refs.bib file
\bibliographystyle{unsrt}
\end{document}